\documentclass[a4paper,11pt]{article}

\usepackage[dvips]{color,graphicx}

\definecolor{red  }{rgb}{1,0,0}
\definecolor{blue }{rgb}{0,0,1}
\definecolor{green}{rgb}{0,1,0}

\newcommand{\vs}[1]{\vspace{#1 mm}}

\renewcommand{\d}{\delta}

\usepackage{graphicx,amssymb,amsmath,bm,latexsym}
\begin{document}

\begin{titlepage}

\vskip .5in

\begin{center}

{\Large\bf  Infinite-dimensional  3-algebra and integrable system } \vskip .5in

{\large Min-Ru Chen$^{a,b,}$\footnote{cmr@henu.edu.cn},
Shi-Kun Wang$^{c,}$\footnote{wsk@amss.ac.cn},
Ke Wu$^{a,}$\footnote{wuke@mail.cnu.edu.cn}
and Wei-Zhong Zhao$^{a,d,}$
\footnote{Corresponding author: zhaowz100@163.com}} \\
\vs{10}
$^a${\em School of Mathematical Sciences, Capital Normal University,
Beijing 100048, China} \\
$^b${\em College of Mathematics and Information Sciences, Henan
University, Kaifeng 475001, China} \\
$^c${\em Institute of Applied Mathematics, Academy of Mathematics and Systems Science, Chinese
Academy of Sciences, Beijing 100190, China} \\
$^d${\em Institute of Mathematics and Interdisciplinary Science,
Capital Normal University, Beijing 100048, China } \\

\vskip .2in \vspace{.3in}

\begin{abstract}
The relation  between the infinite-dimensional  3-algebras and the
dispersionless KdV hierarchy is investigated. Based on the infinite-dimensional  3-algebras,
we derive  two compatible Nambu Hamiltonian structures. Then the dispersionless KdV hierarchy
follows from the Nambu-Poisson evolution equation given the suitable Hamiltonians.
We find that the dispersionless KdV system is not only a bi-Hamiltonian system,
but also a  bi-Nambu-Hamiltonian system. Due to the  Nambu-Poisson evolution equation involving
two Hamiltonians,  more intriguing relationships between these Hamiltonians are  revealed.
As an application, we investigate the system of polytropic gas equations and
derive an integrable gas dynamics system in the framework of Nambu mechanics.

\end{abstract}

\end{center}



\vfill

\end{titlepage}

\section{Introduction}

Infinite-dimensional algebras
have attracted a lot of interest from physical and mathematical
points of view. They have been extensively studied in the past several decades.
In the context of integrable systems, the infinite-dimensional algebras play a very important role.
The Virasoro algebra is an important infinite-dimensional algebra.
The intriguing  relation between the Virasoro algebra and the Korteweg-de Vries (KdV) equation
was found by Gervais and Neveu \cite{Gervais1, Gervais2}.
They considered the equivalent Poisson brackets of the Fourier transform  field
(FTF)  corresponding to the classical  Virasoro algebra and derived
the KdV equation with respect to the FTF from the Poisson evolution equation given a suitable Hamiltonian.
The  similar construction for the super Virasoro algebra and super KdV equation has also
been exploited \cite{Kupershmidt}-\cite{Tanaka2}.
As a higher-spin extension of the
Virasoro algebra, the $W_{N}$ algebra has been found to be
associated to the generalized KdV hierarchies \cite{Mathieu2, Yamagishi}.

To construct the generalized Hamiltonian dynamics, Nambu
\cite{Nambu}  first proposed the notion of 3-bracket. Thus Nambu
dynamics is described by the phase flow given by Nambu-Hamilton
equations of motion which involves two Hamiltonians.
Bayen and Flato  \cite{Bayen} pointed out that in the
classical case Nambu mechanics is equivalent to a
singular Hamiltonian mechanics. The connection between Nambu mechanics
and conventional Hamiltonian ideas has also
been explored by Mukunda and Sudarshan \cite{Mukunda}.
A notion of a Nambu 3-algebra was introduced in
\cite{Takhtajan}  as a natural generalization of a Lie algebra for
higher-order algebraic operations.
Recently a world-volume description of multiple M2-branes using the 3-algebra
was proposed independently by Bagger and Lambert \cite{BL2007}, and Gustavsson \cite{Gustavsson} (BLG).
Since then, great attention has been paid to the BLG theory and 3-algebras \cite{CS2008}-\cite{Chenho}.
The study  of the infinite-dimensional
3-algebras has made big progress in recent years. Curtright et al.
\cite{Curtright} constructed a 3-bracket variant of the
Virasoro-Witt (V-W) algebra  through the use of su(1,1) enveloping
algebra techniques. The Kac-Moody 3-algebra was investigated by Lin
\cite{Lin}.  More recently,  the  $w_{\infty}$ 3-algebra which
satisfies the fundamental identity (FI) condition of 3-algebra was
derived from two different ways.
 Chakrabortty et al.
\cite{Chakrabortty}  gave a $w_{\infty}$ 3-algebra by applying a double scaling
limits on the generators of the $W_{\infty}$ algebra. Chen et al. \cite{Chen}
started from the investigation of the high-order V-W
3-algebra where the parameter $\hbar$ is introduced in the
generators.  By requiring  the  classical limit $\{\  ,\  ,\
\}=\lim\limits_{\hbar \rightarrow 0 }\frac{1}{i\hbar}[\ , \ , \ ]$
to hold , the $w_{\infty}$ 3-algebra follows from the high-order
V-W 3-algebra.

We have  mentioned above the role of infinite-dimensional algebras
in the integrable systems.  An interesting
open question is whether there is a relation between the infinite-dimensional
3-algebras and the integrable equations.
The goal of this paper is to  establish this kind of relation and apply it to the physics system.
We shall  present  two new infinite-dimensional  3-algebras which are the $SDiff(T^2)$ and classical Heisenberg
3-algebras  and show how
these two  3-algebras and already known  $w_{\infty}$ 3-algebra
are related to the dispersionless KdV
hierarchy.  More intriguing properties related to the dispersionless KdV
hierarchy will be revealed.

This paper is organized as follows. In section 2, we investigate
 the infinite-dimensional 3-algebras. In section 3,
we discuss the relation between the
infinite-dimensional algebras and the dispersionless KdV hierarchy in the framework of
Hamiltonian mechanics.  Nambu mechanics is a generalization of classical Hamiltonian mechanics.
In section 4, we establish the relation between the infinite-dimensional
3-algebras and the dispersionless KdV system in the framework of Nambu mechanics.
We point out that  the dispersionless KdV system is
not only a bi-Hamiltonian system, but also a  bi-Nambu-Hamiltonian system.
An application in the gas dynamics system is given in section 5.
We end this paper with
the concluding remarks in section 6.

\vskip5cm

\section{  Infinite-dimensional 3-algebras}

The Nambu mechanics \cite{Nambu} is a generalization of classical Hamiltonian
mechanics, where the Nambu 3-bracket is defined for a triple of
classical observables on the three-dimensional phase space
$\mathbb{R}^3$ with coordinates $\hat x, \hat y, \hat z$ by the
following formula:
\begin{eqnarray}\label{eq:np1}
&&\{f, g, h\}
=\frac{\partial (f, g, h)}{\partial (\hat x,\hat y,\hat z)}=\frac{\partial f}{\partial \hat x}
(\frac{\partial g}{\partial \hat y}\frac{\partial h}{\partial \hat z}-\frac{\partial h}{\partial \hat y}\frac{\partial g}{\partial \hat z})\nonumber\\
&&+\frac{\partial g}{\partial \hat x}
(\frac{\partial h}{\partial \hat y}\frac{\partial f}{\partial \hat z}-\frac{\partial f}{\partial \hat y}\frac{\partial h}{\partial \hat z})
+\frac{\partial h}{\partial \hat x}
(\frac{\partial f}{\partial \hat y}\frac{\partial g}{\partial \hat z}-\frac{\partial g}{\partial \hat y}\frac{\partial f}{\partial \hat z}),
\end{eqnarray}
which satisfies the properties of
skewsymmetry, the Leibniz rule  and the following FI condition:
\begin{eqnarray}\label{eq:np2}
&&\{\{f_1, f_2, f_3\}, f_4, f_5\}+\{f_3, \{f_1, f_2, f_4\}, f_5\}\nonumber\\
&&+\{f_3, f_4, \{f_1, f_2, f_5\}\}=\{f_1, f_2, \{f_3, f_4, f_5\}\}.
\end{eqnarray}

The generators of  $w_{\infty}$
3-algebra are given by
\begin{eqnarray}\label{eq:genw3}
L_m^r=\sqrt{\hat z}\exp[(r-\frac{1}{2})\hat x-2m\hat y],
\end{eqnarray}
where $m,r\in \mathbb{Z}$. The generators (\ref{eq:genw3}) can be
identified with the modes of the deformations of 2-torus \cite{Chakrabortty}.
Substituting the generators (\ref{eq:genw3})
into (\ref{eq:np1}), we obtain the $w_{\infty}$ 3-algebra
\begin{eqnarray}\label{eq:3alg2}
\{L_m^r, L_n^s, L_k^h\}=(h(n-m)+s(m-k)
+r(k-n))L_{m+n+k}^{r+s+h-1}.
\end{eqnarray}

It is well-known  that the Nambu bracket structure of order $n$ on phase space induces infinite
family of subordinated Nambu structures of orders $n-1$ and lower, including
the family of Poisson structures  \cite{Takhtajan}.
Let us define the Poisson bracket
parametrized by $L_0^0$ as
$\{L_m^r, L_n^s\}_{L_0^0}=\{L_m^r, L_n^s, L_0^0\}$.
This parametrized Poisson bracket
satisfies the Jacobi identity, it can be regarded as the generalized $w_\infty$ algebra
\begin{eqnarray}\label{eq:walg}
\{L_m^r, L_n^s\}=(ms-nr)L_{m+n}^{r+s-1},
\end{eqnarray}
where $m,n,r,s\in \mathbb{Z}$. When $r=s=1$,  (\ref{eq:walg}) leads to
the well-known V-W algebra
\begin{eqnarray}\label{eq:vwalg}
\{L_m , L_n \}=(m-n)L_{m+n}.
\end{eqnarray}
It should be pointed
out that the $w_{\infty}$ algebra \cite{Pope} is equivalent to the
algebra of smooth area-preserving diffeomorphisms of the cylinder
$S^1\times R^1$. Its Poisson bracket algebra is also given by
(\ref{eq:walg}), but the superscripts $r,s$ in generators are the
conformal spin with $r,s\geq 1$.

Let us take
\begin{eqnarray}\label{eq:genw4}
T_m^r=\sqrt{\hat z}\exp[\sqrt{2}m\hat x+\sqrt{2}r\hat y],
\end{eqnarray}
where $m, r\in \mathbb{Z}$.
Substituting the generators (\ref{eq:genw4})
into (\ref{eq:np1}), we obtain the following infinite-dimensional 3-algebra:
\begin{eqnarray}\label{eq:3alg4}
\{T_m^r, T_n^s, T_k^h\}=(h(n-m)+s(m-k)+r(k-n))T_{m+n+k}^{r+s+h}.
\end{eqnarray}
From (\ref{eq:3alg4}),  we have
\begin{eqnarray}\label{eq:3alg5}
\{T_m^r, T_n^s\}_{T_0^0}=\{T_m^r, T_n^s, T_0^0\}=(ms-nr)T_{m+n}^{r+s}.
\end{eqnarray}
Note that (\ref{eq:3alg5}) is nothing but a subalgebra of the centerless algebra
of diffeomorphisms of the torus \cite{Iliopoulos, Antoniadis}.
Thus we call (\ref{eq:3alg4}) the $SDiff(T^2)$ 3-algebra.

\section{ Dispersionless KdV system}

The dispersionless KdV system is an important integrable model
which has many applications in physics, such as
unstable fingering patterns of two-dimensional flows of viscous fluids  \cite{Teodorescu},
dynamics of interacting cold atomic gases \cite{Kulkarni} and polytropic gas dynamics   \cite{Popowicz}.
The infinite number of conserved charges $\hat H_n$ of the dispersionless  KdV system
\cite{Das, Brunelli} are given by
\begin{eqnarray}\label{eq:Hkdv}
\hat H_n=\frac{(2n-3)!!}{(2n-2)!!n}\int_0^{2\pi} u^n(x)dx,
\end{eqnarray}
where $n\in \mathbb{Z}_+$ and $(-1)!!=0!!=1$.

Let us introduce the FTF as follows:
\begin{eqnarray}
u(x)=\frac{-i}{2\pi}\sum_{m=-\infty}^{\infty}L_m e^{-imx},
\end{eqnarray}
where $i=\sqrt{-1}$, the field $u(x)$ is periodic in $x$
with period $2\pi$ and has continuous $x$ derivative.
Thus the Poisson bracket realization of the V-W algebra (\ref{eq:vwalg}) can be expressed as
\begin{eqnarray}\label{eq:uu}
\{u(x), u(y)\}_2=u_x\delta(x-y)+2u\delta_x(x-y),
\end{eqnarray}
where the subscript $x$ denotes the derivative with respect to the
variable $x$. By means of (\ref{eq:uu}), it can be shown that the
Hamiltonians (\ref{eq:Hkdv}) are in involution, i.e.,
$\{\hat H_n, \hat H_m\}_2=0$.
Substituting the  Hamiltonians (\ref{eq:Hkdv}) into
the Poisson evolution equation
\begin{eqnarray}\label{eq:2PEE}
\frac{\partial u(t, x)}{\partial t}=\{u, \hat H_n\}
\end{eqnarray}
and using the Poisson bracket (\ref{eq:uu}),
we obtain the dispersionless KdV hierarchy
\begin{eqnarray}\label{eq:DKDVe2}
\frac{\partial u(t, x) }{\partial
 t}&=&A_{n-1}u^{n-1}u_{x}\nonumber\\
&=&\left\{\begin{array}{ccccccccc}
 u_{x}, &n=1\\
\\
\frac{3}{2} u u_{x}, &n=2\\
\\
\frac{15}{8} u^2 u_{x}, &n=3\\
\\
\frac{35}{16} u^3 u_{x}, &n=4\\
\\
\frac{315}{128} u^4 u_{x}, &n=5\\
\\
\vdots &, \\
\end{array}\right.
\end{eqnarray}
where $A_{n}$ is given by
\begin{eqnarray}\label{eq:An}
A_{n}=\frac {(2n+1)!!}{(2n)!!}.
\end{eqnarray}

It has been well known that integrability of many
systems is closely related to their bi-Hamiltonian property \cite{Magri}.
The dispersionless KdV system is a bi-Hamiltonian system. One can associate it to the two kinds of
Poisson brackets called the first and second
Hamiltonian structures. We note that the second Hamiltonian structure (\ref{eq:uu}) of the dispersionless KdV system
can be derived from the classical V-W algebra.  To give its first Hamiltonian structure,
let us turn to another infinite-dimensional algebra, i.e.,  Heisenberg algebra.
Its Poisson bracket realization \cite{Awata} is  given by
\begin{eqnarray}\label{eq:Halg}
\{a_m, a_n\}=m\delta_{m+n, 0}.
\end{eqnarray}
Taking $u(x)=\frac{1-i}{\sqrt{2\pi}}\sum_{m=-\infty}^{+\infty}a_m e^{-imx}$,
then we can rewrite (\ref{eq:Halg})  as
\begin{equation}\label{eq:Halgu}
\{u(x), u(y)\}_1=2\delta_x(x-y).
\end{equation}
It is worth noting that (\ref{eq:Halgu}) is the first  Hamiltonian structure
of the dispersionless KdV system  \cite{Das}.
Based on (\ref{eq:Halgu}),  it is not difficult to derive the dispersionless KdV  hierarchy
(\ref{eq:DKDVe2}) from the Poisson evolution equation (\ref{eq:2PEE}).
By means of (\ref{eq:uu}) and
(\ref{eq:Halgu}), we obtain the ratio between $\{u, \hat{H}_m\}_2$ and $\{u, \hat{H}_m\}_1$ as follows:
\begin{eqnarray}\label{eq:ratio12}
\frac{\{u, \hat{H}_m\}_2}{\{u, \hat{H}_m\}_1}=\frac{2m-1}{2m-2}u.
\end{eqnarray}

\section{Infinite-dimensional 3-algebras and Dispersionless KdV system}

In the previous section, we have investigated the relation  between the
infinite-dimensional algebras  and  the dispersionless KdV system.
We turn our attention back to the case of the infinite-dimensional  3-algebras.
Let us introduce the FTF  $u(x, y)$ as follows:
\begin{eqnarray}\label{eq:ftf1}
 u(x, y)=\frac{1}{4\pi^2}\sum_{m, r}L_m^r
e^{-imx-iry}.
\end{eqnarray}
It is assumed that $u$ has continuous $x$ and $y$ derivatives and
satisfies the periodic boundary condition $u(t, x+2k\pi,
y+2l\pi)=u(t, x, y),(k, l\in\mathbb{Z})$.  Using
(\ref{eq:ftf1}), we can rewrite $w_{\infty}$ 3-algebra (\ref{eq:3alg2}) as  the following
Nambu 3-bracket relation of the FTF:
\begin{eqnarray}\label{eq:3algftf1}
&&\{u(x_1, y_1), u(x_2, y_2), u(x_3, y_3)\}_2\nonumber\\
&&=[u_{x_1}(x_1, y_1)\delta(x_1-x_2)\delta(x_1-x_3)\delta_{y_1}(y_1-y_2)\delta(y_1-y_3)-u_{x_1}(x_1, y_1)\delta(x_1-x_2)\nonumber\\
&&\delta(x_1-x_3)\delta(y_1-y_2)\delta_{y_1}(y_1-y_3)+u_{y_1}(x_1, y_1)\delta(x_1-x_2)\delta_{x_1}(x_1-x_3)\delta(y_1-y_2)\nonumber\\
&&\delta(y_1-y_3)-u_{y_1}(x_1,y_1)\delta_{x_1}(x_1-x_2)\delta(x_1-x_3)\delta(y_1-y_2)\delta(y_1-y_3)-3u(x_1, y_1)\nonumber\\
&&\delta_{x_1}(x_1-x_2)\delta(x_1-x_3)\delta(y_1-y_2)\delta_{y_1}(y_1-y_3)+3u(x_1, y_1)\delta(x_1-x_2)\delta_{x_1}(x_1-x_3)\nonumber\\
&&\delta_{y_1}(y_1-y_2)\delta(y_1-y_3)-iu(x_1, y_1)\delta(x_1-x_2)\delta_{x_1}(x_1-x_3)\delta(y_1-y_2)\delta(y_1-y_3)\nonumber\\
&&+iu(x_1, y_1)\delta_{x_1}(x_1-x_2)\delta(x_1-x_3)\delta(y_1-y_2)\delta(y_1-y_3)]e^{-iy_1}.
\end{eqnarray}
where the subscripts $x_l$ and $y_l$ throughout this paper
denote the partial derivatives  with respect to the variables $x_l$
and $y_l$, respectively.

Let us consider the
following extended Hamiltonians  of (\ref{eq:Hkdv}) with respect to
the variable $y$:
\begin{eqnarray}\label{eq:H3}
&&H_n=\frac{(2n-3)!!}{(2n-2)!!n}\int\limits_{0}^{2\pi}\int\limits_{0}^{2\pi}
u^n(x, y)dxdy,
\end{eqnarray}
where $u(x, y)=\hat u(x)v(y)$.

By means of (\ref{eq:3algftf1}), we find that the Hamiltonians (\ref{eq:H3})
are in involution with the Nambu
3-bracket structure, i.e.,  $\{H_k, H_m, H_n\}_2=0$.
In the framework of Nambu mechanics, the  Nambu-Hamilton equations of motion
involve two Hamiltonians.  Let us consider  the following Nambu-Poisson (N-P) evolution equation:
\begin{eqnarray}\label{eq:hmhn}
\frac{\partial {u(t, x, y)}}{\partial t}=\{u,
H_m, H_n\},
\end{eqnarray}
where $m, n\in \mathbb{Z}_{+}$ and $m\neq n$. Substituting the
Hamiltonians (\ref{eq:H3}) into (\ref{eq:hmhn}) and using the Nambu 3-bracket relation
(\ref{eq:3algftf1}), we obtain
\begin{eqnarray}\label{eq:hnhmr}
\frac{\partial u(t, x, y)}{\partial t}&=&\{u, H_m, H_n\}_2
=S_{mn}A_{m+n-2}u^{m+n-2}u_{x},
\end{eqnarray}
where
$S_{mn}=\frac{(m-n)(2m-3)!!(2n-3)!!(2m+2n-4)!!}{(2m-2)!!(2n-2)!!(2m+2n-3)!!}ie^{-iy}$.
Let us take $\hat{t}_{mn}=S_{mn}v^{m+n-2}(y)t$ and require
$\frac{\partial v(y)}{\partial t}=0$, we obtain the dispersionless
KdV hierarchy
\begin{eqnarray}\label{eq:DKDVe3}
\frac{\partial \hat{u}(t, x)}{\partial
\hat{t}_{mn}}=A_{m+n-2}\hat{u}^{m+n-2}\hat{u}_{x}
\end{eqnarray}
from (\ref{eq:hnhmr}), where $m+n\ge 3$. Note that (\ref{eq:DKDVe3})
matches with the dispersionless  KdV hierarchy (\ref{eq:DKDVe2})
except for the first equation in (\ref{eq:DKDVe2}).

Let us consider the FTF (\ref{eq:ftf1}) with substitution of the generators $T^r_m$ of
$SDiff(T^2)$ 3-algebra for  $L^r_m$ of $w_{\infty}$ 3-algebra, it is easy to rewrite
the $SDiff(T^2)$ 3-algebra as the Nambu 3-bracket structure of the FTF $u$.
Then based on this equivalent Nambu 3-bracket relation,
after a straightforward calculation, we find $\frac{\partial u(t,x,y)}{\partial t}=\{u, H_m, H_n\}=0$.

The central extensions of the infinite-dimensional algebras have been well investigated.
The non-trivial central extensions of various  infinite-dimensional algebras have been determined, such as
$w_{\infty}$ and  $W_{\infty}$ algebras  \cite{Pope}  and $SDiff(T^2)$ algebra \cite{Iliopoulos, Antoniadis}.
An  open question is whether there exist  the non-trivial central extensions for
the infinite-dimensional 3-algebras.
Even though we do not know the non-trivial central extension of $SDiff(T^2)$ 3-algebra (\ref{eq:3alg4}),
nevertheless it still admits the trivial central extensions.

Let us consider  the $SDiff(T^2)$ 3-algebra (\ref{eq:3alg4}) with the following trivial central extension:
\begin{eqnarray}\label{eq:ce3alg4}
\{T_m^r, T_n^s, T_k^h\}&=&[h(n-m)+s(m-k)+r(k-n)](T_{m+n+k}^{r+s+h}+4\pi^2\d_{m+n+k, 0}\d_{r+s+h, 1})\nonumber\\
&=&[h(n-m)+s(m-k)+r(k-n)]T_{m+n+k}^{r+s+h}\nonumber\\
&&+4\pi^2[3(ms-nr)+n-m]\d_{m+n+k, 0}\d_{r+s+h, 1}.
\end{eqnarray}
It is easy to verify that (\ref{eq:ce3alg4}) satisfies the FI condition (\ref{eq:np2}) and skewsymmetry.
By means of the FTF
$u(x, y)=\frac{1}{4\pi^2}\sum_{m, r}T_m^r e^{-imx-iry}$,
we may  rewrite (\ref{eq:ce3alg4}) as
\begin{eqnarray}\label{eq:uc3alg4}
&&\{u(x_1,y_1), u(x_2,y_2), u(x_3,y_3)\}_1\nonumber\\
&=&[u_{x_1}(x_1,y_1)\delta(x_1-x_2)\delta(x_1-x_3)\delta_{y_1}(y_1-y_2)\delta(y_1-y_3)-u_{x_1}(x_1,y_1)\delta(x_1-x_2)\nonumber\\
&&\delta(x_1-x_3)\delta(y_1-y_2)\delta_{y_1}(y_1-y_3)+u_{y_1}(x_1,y_1)\delta(x_1-x_2)\delta_{x_1}(x_1-x_3)\delta(y_1-y_2)\nonumber\\
&&\delta(y_1-y_3)-u_{y_1}(x_1,y_1)\delta_{x_1}(x_1-x_2)\delta(x_1-x_3)\delta(y_1-y_2)\delta(y_1-y_3)+3u(x_1,y_1)\nonumber\\
&&\delta(x_1-x_2)\delta_{x_1}(x_1-x_3)\delta_{y_1}(y_1-y_2)\delta(y_1-y_3)-3u(x_1,y_1)\delta_{x_1}(x_1-x_2)\delta(x_1-x_3)\nonumber\\
&&\delta(y_1-y_2)\delta_{y_1}(y_1-y_3)]+A,
\end{eqnarray}
where $A$ is the contribution of the trivial central extension term and given by
\begin{eqnarray}\label{eq:A}
A&=&e^{-iy_1}[3\delta (x_1-x_2)\delta _{x_1}(x_1-x_3)\delta _{y_1}(y_1-y_2)\delta (y_1-y_3)-3\delta _{x_1}(x_1-x_2)\nonumber\\
&&\delta (x_1-x_3)\delta (y_1-y_2)\delta _{y_1}(y_1-y_3)+i\delta _{x_1}(x_1-x_2)\delta (x_1-x_3)\delta (y_1-y_2)\nonumber\\
&&\delta (y_1-y_3)-i\delta (x_1-x_2)\delta _{x_1}(x_1-x_3)\delta (y_1-y_2)\delta (y_1-y_3)].
\end{eqnarray}

Due to the contribution of this trivial central extension term,
the N-P evolution equation of the FTF $u$ becomes
\begin{eqnarray}\label{eq:s2dkdv}
\frac{\partial u(t,x,y)}{\partial t}
&=&\{u, H_m, H_n\}_1
=\widetilde{S}_{mn}A_{m+n-3}u^{m+n-3}u_{x},
\end{eqnarray}
where $\widetilde{S}_{mn}=\frac{(m-n)(2m-3)!!(2n-3)!!(2m+2n-6)!!}{(2m-2)!!(2n-2)!!(2m+2n-5)!!}ie^{-iy}$.
By taking  $\hat{t}_{mn}={\widetilde S}_{mn}v^{m+n-3}(y)t$ and
$\frac{\partial v(y)}{\partial t}=0$, (\ref{eq:s2dkdv}) leads  the dispersionless
KdV hierarchy (\ref{eq:DKDVe3}).
We also observe that the Hamiltonians (\ref{eq:H3})
satisfy $\{H_k, H_m, H_n\}_1=0.$

After understanding the crucial importance of this trivial central extension,
we may only consider its contribution. Thus we have
\begin{eqnarray}\label{Heisenberd3alg}
&&\{u(x_1, y_1), u(x_2, y_2), u(x_3, y_3)\}_1\nonumber\\
&&=e^{-iy_1}[3\delta (x_1-x_2)\delta _{x_1}(x_1-x_3)\delta _{y_1}(y_1-y_2)\delta (y_1-y_3)-3\delta _{x_1}(x_1-x_2)\nonumber\\
&&\delta (x_1-x_3)\delta (y_1-y_2)\delta _{y_1}(y_1-y_3)+i\delta _{x_1}(x_1-x_2)\delta (x_1-x_3)\delta (y_1-y_2)\nonumber\\
&&\delta (y_1-y_3)-i\delta (x_1-x_2)\delta _{x_1}(x_1-x_3)\delta (y_1-y_2)\delta (y_1-y_3)].
\end{eqnarray}
Based on (\ref{Heisenberd3alg}), we may directly derive  (\ref{eq:s2dkdv}).
Comparing  (\ref{eq:hnhmr})  with  (\ref{eq:s2dkdv}), we have
$\frac{\{u, H_m, H_n\}_2}{\{u, H_m, H_n\}_1}=u$.
The form of this ratio is the same as (\ref{eq:ratio12}) except for the
coefficient.

Introducing $u(x, y)=\frac{1}{\sqrt[3]{4\pi^4}}\sum_{m, r}a_m^r e^{-imx-iry}$,
we obtain the  following 3-algebra  from (\ref{Heisenberd3alg}):
\begin{eqnarray}\label{eq:Heisenberg}
\{a_m^r, a_n^s, a_k^h\}&=&\frac{1}{4}[3(ms-nr)+(n-m)]\delta _{m+n+k, 0}\delta _{r+s+h, 1}.
\end{eqnarray}
When $r=s=1$,  $k=0$ and $h=-1$, (\ref{eq:Heisenberg}) reduces to the classical Heisenberg algebra (\ref{eq:Halg}).
For this reason we call (\ref{eq:Heisenberg}) the classical  Heisenberg 3-algebra.

We have associated the dispersionless KdV system  to the two kinds of Nambu 3-bracket which are
(\ref{Heisenberd3alg}) and (\ref{eq:3algftf1})
called the first and second Nambu Hamiltonian structures, respectively.
These two  Nambu Hamiltonian structures  can be derived from the classical Heisenberg and $w_{\infty}$
3-algebras, respectively. It can be very easily observed that their sum is again a Nambu Hamiltonian structure.
It implies that these two Nambu Hamiltonian structures are compatible.
Thus the dispersionless KdV system is not only a bi-Hamiltonian system, but also a  bi-Nambu-Hamiltonian system.

For the Hamiltonians (\ref{eq:H3}), we note that they are in involution
with the first and second Nambu Hamiltonian structures, respectively.
To achieve a better understanding of the relationship between these Hamiltonians, let us consider
the Hamiltonian pairs $(H_n, H_m)$ and $(H_k, H_l)$ with $n+m=k+l$. We find that
the ratios between $\{u(x, y), H_m, H_n\}_i$ and $\{u(x, y), H_k, H_l\}_i$, i=1,2,  are
\begin{eqnarray}\label{eq:ratio3}
\frac{\{u(x, y), H_m, H_n\}_i}{\{u(x, y), H_k, H_l\}_i}
=\phi\frac{2m-k-l}{k-l},
\end{eqnarray}
where $\phi$ is given by the hypergeometry function
\begin{eqnarray}
\phi={_5F_4}\left[\begin{array}{ccccc}
\frac{1}{2} m+\frac{1}{2}, &m-1, &m-k, &m-l, &\frac{1}{2}\\
\\
\frac{1}{2} m-\frac{1}{2}, &k,  &l,  &m-\frac{1}{2}&
\end{array}\right].
\end{eqnarray}
Although the ratios (\ref{eq:ratio3}) are not equal to one, we note
that the corresponding N-P evolution equation
may lead to the same dispersionless  KdV equation with
the appropriate rescaling. It implies that there is an intrinsic
equivalent relation between these Hamiltonian pairs.
When $s=m+n$ is even and odd for the Hamiltonian pairs $(H_m, H_n)$,
for a given Nambu Hamiltonian structure,  there
are $\frac{s-2}{2}$ and $\frac{s-1}{2}$ different Hamiltonian pairs
which may derive to  the same dispersionless  KdV equation from
the N-P evolution equation, respectively.
This can be regarded as the generic property of the  dispersionless  KdV hierarchy
from the viewpoint of Nambu mechanics.

The Nambu 3-bracket can also be generalized to the bracket of n functions
$f_i = f_i(\hat x_1, \hat x_2, \cdots, \hat x_n)$ defined by \cite{Takhtajan}
\begin{eqnarray}\label{eq:nbracket}
\{f_1, f_2, \cdots, f_n\}=\frac{\partial(f_1, f_2, \cdots, f_n)}{\partial(\hat
x_1, \hat x_2, \cdots, \hat x_n)}.
\end{eqnarray}
Substituting the generators (\ref{eq:genw3}) and (\ref{eq:genw4})
into (\ref{eq:nbracket}), we obtain the $w_{\infty}$ and $SDiff(T^2)$ n-algebras as follows:
\begin{eqnarray}
\{L_{m_1}^{r_1}, L_{m_2}^{r_2}, \cdots, L_{m_n}^{r_n}\}=
\{T_{m_1}^{r_1}, T_{m_2}^{r_2}, \cdots, T_{m_n}^{r_n}\}
=0,
\end{eqnarray}
where $n\ge 4$.
Due to these two null n-algebras, it is not difficult to verify that the Hamiltonians (\ref{eq:H3})
satisfy
\begin{eqnarray}
\{H_{i_1}, H_{i_2}, \cdots, H_{i_n}\}=0,
\end{eqnarray}
and the generalized N-P evolution equation is
\begin{eqnarray}
\frac{\partial u(t, x, y)}{\partial t}=\{u, H_{i_1}, H_{i_2}, \cdots,  H_{i_{n-1}}\}=0.
\end{eqnarray}

\section{Application}

As an application, let us consider the $2+1$ dimension isentropic polytropic gas dynamics equation \cite{Whitham}
\begin{eqnarray}\label{eq:Euler}
&&\frac{\partial \rho}{\partial t}+(\rho u)_x+(\rho v)_y=0,\nonumber\\
&&\frac{\partial u}{\partial t}+u\frac{\partial u}{\partial x}+v\frac{\partial u}{\partial y}+k\gamma\rho^{\gamma-2}\rho_x=F_1,\nonumber\\
&&\frac{\partial v}{\partial t}+u\frac{\partial v}{\partial x}+v\frac{\partial v}{\partial y}+k\gamma\rho^{\gamma-2}\rho_y=F_2,
\end{eqnarray}
where $\rho(t,x,y)$ is the gas density,  $(u,v)$ is the velocity vector of gas flow, $(F_1, F_2)$ is the  body force vector per unit mass,
$\gamma$ and $k$ are the constants, $\gamma$ denotes the ratio of the specific heats.
The system of polytropic gas equations has attracted a great deal of attention due to its wide applications in physics.
Let us focus on a  special gas motion with $\gamma=3$, $v=0$, $F_2=3k\rho\rho_y$ and $F_1=0$, thus (\ref{eq:Euler}) becomes
\begin{eqnarray}\label{eq:uvt}
\frac{\partial \rho(t,x,y)}{\partial t}&=&
-(\rho u)_{x},\nonumber\\
\frac{\partial u(t,x,y)}{\partial t}&=&
-\frac{1}{2}(u^2+3k\rho^2)_{x}.
\end{eqnarray}
Under the dimensional reduction of $y=0$, (\ref{eq:uvt}) reduces to the well-investigated  gas dynamics equation in $1+1$ dimensions,
which admits two infinite sets of conserved charges , i.e.,  ¡°Eulerian¡± and ¡°Lagrangian¡±
conserved charges, respectively \cite{Gumral, Olver}.

Let us take
\begin{eqnarray}\label{eq:urho}
u&=&-\frac{1}{4}ie^{-iy}(A_{+}+A_{-}),\nonumber\\
\rho&=&-\frac{1}{4\sqrt{3k}}ie^{-iy}(A_{+}-A_{-}).
\end{eqnarray}
Thus (\ref{eq:uvt}) can be rewritten as
\begin{eqnarray}\label{eq:AB1}
\frac{\partial A_{\pm}(t,x,y)}{\partial t}=
\frac{i}{2}e^{-iy}A_{\pm} A_{\pm x}.
\end{eqnarray}
We immediately recognize that (\ref{eq:AB1}) is nothing but the
first equation of the hierarchy  (\ref{eq:hnhmr}).
It implies that we may derive (\ref{eq:uvt}) in the framework of Nambu mechanics.
To achieve this result, we take the Hamiltonians
\begin{eqnarray}\label{eq:Hamiltonian}
&&H_{\pm n}=\frac{(2n-3)!!}{(2n-2)!!n}\int\limits_{0}^{2\pi}\int\limits_{0}^{2\pi}
A_{\pm}^n(x,y)dxdy\nonumber\\
&&=\frac{(2n-3)!!}{(2n-2)!!n}\int\limits_{0}^{2\pi}\int\limits_{0}^{2\pi}
(2i)^ne^{iny}(u\pm\sqrt{3k}\rho)^ndxdy.
\end{eqnarray}
When $y=0$, the Hamiltonians (\ref{eq:Hamiltonian}) are the combination of the ¡°Eulerian¡± and ¡°Lagrangian¡±
conserved charges of $1+1$ dimensional gas dynamics equation.

Not as the Poisson evolution equation, the N-P evolution equation involves
two Hamiltonians.  Let us focus on the Hamiltonian pairs $(H_{\pm 1}, H_{\pm 2})$ and $(H_{\pm 1}, H_{\pm 3})$.
Substituting these two Hamiltonian pairs into the N-P evolution equation and using
the Nambu Hamiltonian structures (\ref{eq:3algftf1}) and (\ref{Heisenberd3alg})
with substitution of $A_{\pm}(x,y)$ for $u(x,y)$, respectively,
we may derive (\ref{eq:AB1}).
From applying (\ref{eq:3algftf1}) and (\ref{Heisenberd3alg}),
we see that the classical Heisenberg and $w_{\infty}$
3-algebras play an crucial role in the derivation of (\ref{eq:AB1}).
Moreover the deep insight into the relationship between $H_{\pm i}$, $i=1,2,3,$
has been achieved.
Since  the isentropic gas dynamics
equation (\ref{eq:uvt}) can be expressed as the two non-interacting dispersionless KdV
equations (\ref{eq:AB1}),  its solution can be obtained through the implicit  solution of (\ref{eq:AB1})
$A_{\pm}(t,x,y)=\phi_{\pm}(\frac{1}{2}ie^{-iy}A_{\pm}t+x)\psi_{\pm}(y)$, where  $\phi_{\pm}$ and $\psi_{\pm}$ are the arbitrary differentiable functions.
Due to no dispersive term in (\ref{eq:AB1}),  it can be observed a transition of wave shape
from conservative to dissipative behaviour.
A remarkable feature of the system (\ref{eq:uvt}) is that
the body force $F_2$ should depend on the gas density such that the velocity along the $y$ direction is zero.

\section{Concluding Remarks}

We have investigated the relation between the
infinite-dimensional  3-algebras and the dispersionless KdV hierarchy.
By introducing the FTFs, we rewrote the classical Heisenberg and $w_{\infty}$
3-algebras as the first and second
Nambu 3-bracket structures of the FTFs, respectively.
By choosing the suitable extended Hamiltonians of the dispersionless KdV system, we found
that these  Hamiltonians are in involution for these two
Nambu 3-bracket structures. The dispersionless KdV hierarchy follows from the N-P
evolution equation with these Hamiltonians.
Note that  these two Nambu Hamiltonian structures are compatible. Thus the dispersionless KdV system is
not only a bi-Hamiltonian system, but also a  bi-Nambu-Hamiltonian system.
Due to the  N-P evolution equation involving  two Hamiltonians,   more intriguing
relationships between these Hamiltonians  have been revealed.
As an application of the infinite-dimensional  3-algebras, we
derived an integrable generalized gas dynamics system in the framework of Nambu mechanics.
It should be pointed out that we just deal with a simple nonlinear evolution equation.
For the future investigations, it is of interest to derive other integrable nonlinear evolution equations
related with the infinite-dimensional 3-algebras. Thus  more applications of the infinite-dimensional 3-algebras
in physics will be possible. Furthermore  the connection  between Nambu mechanics and
Hamiltonian mechanics \cite{Bayen, Mukunda} deserves further study with the help of the infinite-dimensional 3-algebras.
We believe that the infinite-dimensional  3-algebras may lead to a better understanding of the integrable systems.
This may shed new light on the integrable systems and Nambu mechanics.

\section*{Acknowledgements}

The authors are indebted to Prof. Y.K. Lau for valuable discussion.
We would like to thank the referees for their helpful comments.
This work is partially supported by NSF projects
(10975102 and 11031005), KZ201210028032 and
PHR201007107.



\begin{thebibliography}{21}


\bibitem{Gervais1} J.L. Gervais and A. Neveu,
Dual string spectrum in Polyakov's quantization (II). Mode separation,
Nucl. Phys. B {\bf 209} (1982) 125.
\bibitem{Gervais2} J.L. Gervais, Infinite family of polynomial functions of the Virasoro
generators with vanishing Poisson brackets,  Phys. Lett. B {\bf 160} (1985) 277.
\bibitem{Kupershmidt} B. A. Kupershmidt, A super Korteweg-de Vries equation: An integrable system,
Phys. Lett. A {\bf 102} (1984) 213.
\bibitem{CK} M. Chaichian and P. Kulish, Superconformal algebras and their relation to
integrable nonlinear systems, Phys. Lett. B {\bf 183} (1987) 169.
\bibitem{BG} A. Bilal and J.L. Gervais, Superconformal algebra and super-kdv equation:
two infinite families of polynomial functions with vanishing poisson brackets,
Phys. Lett. B {\bf 211} (1988) 95.
\bibitem{Kim} Q. Ho-Kim, (Super) Korteveg ¡ªde Vries equation as a (super)
conformal field theory, Phys. Rev. D {\bf 36} (1987) 3829.
\bibitem{Mathieu} P. Mathieu, Superconformal algebra and supersymmetric Korteweg-de Vries equation,
Phys. Lett. B {\bf 203} (1988) 287.
\bibitem{Tanaka2} K. Tanaka,  Solvable two-dimensional supersymmetric models and the supersymmetric Virasoro algebra,
Phys. Rev. D {\bf 42} (1990) 2745.
\bibitem{Mathieu2} P. Mathieu, Extended classical conformal algebras and the second hamiltonian
structure of Lax equations, Phys. Lett. B {\bf 208} (1988) 101.
\bibitem{Yamagishi} K. Yamagishi, The KP hierarchy and extended Virasoro algebras,
Phys. Lett. B {\bf 205} (1988) 466.
\bibitem{Nambu}
Y. Nambu, Generalized Hamiltonian dynamics, Phys. Rev. D {\bf 7} (1973) 2405.
\bibitem{Bayen}
F. Bayen and M. Flato, Remarks concerning Nambu's generalized mechanics,
Phys. Rev. D {\bf 11} (1975) 3049.
\bibitem{Mukunda}
N. Mukunda and E.C.G. Sudarshan, Relation between Nambu and Hamiltonian mechanics,
Phys. Rev. D {\bf 13} (1976) 2846.
\bibitem{Takhtajan}
 L. Takhtajan, On foundation of the generalized Nambu mechanics, Commun. Math. Phys. {\bf 160}
 (1994) 295 [hep-th/9301111].
\bibitem{BL2007}
J. Bagger and N. Lambert, Modeling multiple M2's, Phys. Rev. D {\bf 75} (2007) 045020
[hep-th/0611108];  Gauge symmetry and supersymmetry of multiple M2-branes,
Phys. Rev. D {\bf 77} (2008) 065008 [arXiv:0711.0955]; Comments on multiple M2-branes,
JHEP {\bf 02} (2008) 105 [arXiv:0712.3738].
\bibitem{Gustavsson}
A. Gustavsson, Algebraic structures on parallel M2-branes, Nucl. Phys. B {\bf 811} (2009) 66
[arXiv:0709.1260].
\bibitem{CS2008}
S.A. Cherkis and C. S\"{a}mann, Multiple M2-branes and Generalized 3-Lie algebras,
Phys. Rev. D {\bf 78} (2008) 066019 [arXiv:0807.0808].
\bibitem{Papadopoulos}
G. Papadopoulos, M2-branes, 3-Lie algebras and Pl\"{u}cker relations, JHEP {\bf 05} (2008) 054
[arXiv:0804.2662].
\bibitem{Mukhi}
S. Mukhi and C. Papageorgakis, M2 to D2, JHEP {\bf 05} (2008) 085 [arXiv:0803.3218].
\bibitem{Raamsdonk}
M. Van Raamsdonk, Comments on the Bagger-Lambert theory and multiple M2-branes,
JHEP {\bf 05} (2008) 105 [arXiv:0803.3803].
\bibitem{Gomis}
J. Gomis, G. Milanesi and J.G. Russo, Bagger-Lambert theory for general Lie algebras, JHEP
{\bf 06} (2008) 075 [arXiv:0805.1012].






\bibitem{Ho1}
P.M. Ho and Y. Matsuo, M5 from M2, JHEP {\bf 06} (2008) 105 [arXiv:0804.3629].
\bibitem{Ho2}
P.M. Ho, Y. Imamura, Y. Matsuo and S. Shiba, M5-brane in three-form flux and multiple
M2-branes, JHEP {\bf 08} (2008) 014 [arXiv:0805.2898].
\bibitem{Chenho}
C.H. Chen, K. Furuuchi, P.M. Ho and T. Takimi, More on the Nambu-Poisson M5-brane theory:
scaling limit, background independence and an all order solution to the Seiberg-Witten map,
JHEP {\bf 10} (2010) 100 [arXiv:1006.5291].
\bibitem{Curtright}
T.L. Curtright, D.B. Fairlie and C.K. Zachos, Ternary Virasoro-Witt algebra, Phys. Lett. B
{\bf 666} (2008) 386 [arXiv:0806.3515].
\bibitem{Lin}
 H. Lin, Kac-Moody extensions of 3-algebras and M2-branes, JHEP {\bf 07} (2008) 136 [arXiv:0805.4003].
\bibitem{Chakrabortty}
S. Chakrabortty, A. Kumar and S. Jain, $w_{\infty}$ 3-algebra, JHEP {\bf 09} (2008) 091 [arXiv:0807.0284].
\bibitem{Chen} M.R. Chen, K. Wu and W.Z. Zhao,
Super $w_{\infty}$ 3-algebra,  JHEP {\bf 09} (2011) 090 [arXiv:1107.3295].
\bibitem{Pope}
C.N. Pope, L.J. Romans and X. Shen, The complete structure of $W_{\infty}$,
Phys. Lett. B {\bf 236} (1990) 173.
\bibitem{Iliopoulos}
E.G. Floratos and J. Iliopoulos, A note on the classical symmetries of the closed bosonic membranes,
Phys. Lett. B {\bf 201} (1988) 237.
\bibitem{Antoniadis}
I. Antoniadis, P. Ditsas, E. Floratos and J. Iliopoulos,
New realizations of the Virasoro algebra as membrane symmetries,
Nucl. Phys. B {\bf 300} (1988) 549.
\bibitem{Teodorescu}
R. Teodorescu, P. Wiegmann and A. Zabrodin,
Unstable Fingering Patterns of Hele-Shaw Flows as a Dispersionless Limit of the Kortweg-de Vries Hierarchy,
Phys. Rev. Lett. {\bf 95} (2005) 044502 [cond-mat/0502179].
\bibitem{Kulkarni}
M. Kulkarni and A.G. Abanov, Hydrodynamics of cold atomic gases in the limit of weak nonlinearity,
dispersion, and dissipation, Phys. Rev.  A {\bf 86} (2012)  033614 [arXiv:1205.5917].
\bibitem{Popowicz}
A. Das and Z. Popowicz, Supersymmetric polytropic gas dynamics,
Phys. Lett. A {\bf 296} (2002) 15 [hep-th/0109223].
\bibitem{Das}
A. Choudhuri, B. Talukdar and U. Das,
Lagrangian Approach to Dispersionless KdV Hierarchy, SIGMA  {\bf 3} (2007) 096 [arXiv:0706.0314].
\bibitem{Brunelli} J.C. Brunelli, Hamiltonian Structures for the Generalized Dispersionless KdV Hierarchy,
Rev. Math. Phys. {\bf 8} (1996) 1041 [solv-int/9601001].
\bibitem{Magri}
F. Magri, A simple model of the integrable Hamiltonian equation, J. Math. Phys. {\bf 19} (1978) 1156.
\bibitem{Awata}
H. Awata, H. Kubo, S. Odake and J. Shiraishi,
Virasoro-type Symmetries in Solvable Models, hep-th/9612233.
\bibitem{Whitham}
G.B. Whitham, Linear and Nonlinear Waves, Wiley, New York (1974).
\bibitem{Gumral}
H. G\"{u}mral and Y. Nutku, MultiHamiltonian structure of equations of hydrodynamic type, J. Math. Phys. {\bf 31} (1990) 2606.
\bibitem{Olver}
P.J. Olver and Y. Nutku, Hamiltonian structures for systems of hyperbolic conservation laws,
J. Math. Phys. {\bf 29} (1988) 1610.



\end{thebibliography}
\end{document}